\documentclass[letter,twocolumn]{jpsj3}
\usepackage{txfonts}

\usepackage{bm}

\title{Phase Transitions for Cuboc Orders in Stacked Kagome Heisenberg Systems}
\author{Kouichi Seki, and Kouichi Okunishi$^1$}
\inst{Graduate School of Science and Technology, Niigata University, Niigata 950-2181, Japan \\
$^1$Department of Physics, Niigata University, Niigata 950-2181, Japan}

\abst{
Using the event-chain Monte Carlo (MC) algorithm,  we investigate phase transitions of the stacked Kagome Heisenberg systems with classical vector spins including up to the 3rd-nearest-neighbor couplings.
In particular, we focus on two types of non-coplanar spin orders ---cuboc1 and cuboc2 orders,  both of which have twelve-sublattice structures accompanying the translational-symmetry breaking.  
We perform event-chain MC simulations up to  $L=72$, where $L$ represents the linear dimension of the stacked Kagome lattice, and then find that the cuboc1 transition shows 2nd-order-transition behaviors with a tendency to a weak-1st-order transition up to $L=72$, while the cuboc2 transition is basically described by the 1st-order transition.
We then discuss the above transitions in connection with the effective Landau-Ginzburg-Wilson theory with the O(3)$\times$O(3) symmetry.
}

\begin{document}
\maketitle

Frustrated spin systems often provide fascinating physics associated  with nontrivial spin orders having non-colinear or non-coplanar spin structures.
Recently, a variety of exotic ground-state spin orders were classified for the Kagome-lattice classical Heisenberg antiferromagnets including up to the 3rd-nearest-neighbor couplings.\cite{Messio}
Among various spin orders for the Kagome Heisenberg systems, we particularly focus on the cuboc orders, which are defined as non-coplanar spin configurations with twelve sublattice structure specified by the triple-${\bm q}$ structures in the momentum space.\cite{Domenge, Domenge2}
Moreover, experiments on CsCrF$_4$\cite{Manaka,Seki}, NaBa$_2$Mn$_3$F$_{11}$\cite{Ishikawa} and  Cu$_3$Zn(OH)$_6$Cl$_2$\cite{Kapella, Kapela_cuboc1} suggest the relevance of the cuboc orders to realistic experimental situations.

Although the planar-lattice spin models with continuous symmetries have no finite-temperature order, the inter-layer coupling may lift the exotic ground-state orders into the finite-temperature orders with the spontaneous symmetry breaking, which provide rich physics associated with the phase transitions and critical behaviors.
Actually, the stacked triangular-lattice Heisenberg antiferromagnets have been extensively studied so far, where the effective  Landau-Ginzburg-Wilson (LGW) theory based on the O(3)$\times$O(2) symmetry plays a central role to understand the universality reflecting the double-${\bm q}$ structure of the planar 120$^\circ$ order.\cite{Kawamura_1,  Kawamura1988, Kawamura_1990,  Azaria, Loison, Kawamura_1998,  Pelissetto_2001, Pelissetto_2002, Itakura, Peles, Delamotte,Calabrese,Thanh}
However, phase transitions associated with the non-coplanar spin orders, which are characterized by the triple-${\bm q}$ structures\cite{Kawamura_1990},  have been less explored, because systematical analysis of the complicated non-coplanar orders is difficult from the numerical simulation point of view. 

A crucial point on the stacked Kagome systems is that they can realize two similar but different types of cuboc orders, which are mentioned as cuboc1\cite{Kapela_cuboc1} and cuboc2\cite{Domenge} in Ref. \citen{Messio}; 
although these two cuboc orders have the same symmetry in the spin space,  the triple-${\bm q}$ wave vectors specifying their Bragg peak points are located at different positions in the momentum space.
Thus, the stacked Kagome systems can be an interesting stage for investigating the universality of the cuboc transitions, which may be described by the O(3)$\times$O(3) LGW theory.
In addition, to clarify the stability of the orders would be essential for experiments on the Kagome lattice compounds.

In this letter, we investigate phase transitions of the stacked Kagome Heisenberg systems depicted in Fig. 1(a), using the event-chain Monte Carlo (MC) algorithm,  which was originally introduced for hard core particle systems\cite{Bernard, Michel1} and recently developed to the classical spin systems\cite{Michel2, Nishikawa}.
The event-chain method is a rejection free MC algorithm satisfying the global-balance condition, and  enable us to achieve the proper relaxation even for the complicated spin orders in frustrated spin systems.
We perform MC simulations for the stacked Kagome systems exhibiting the cuboc orders up to about $10^6$ spins, and  then analyze the natures of the cuboc transitions in detail. 
We also discuss the connection to the O(3)$\times$O(3) LGW theory.

\begin{figure}[tb]
\centering\includegraphics[width=0.8\linewidth]{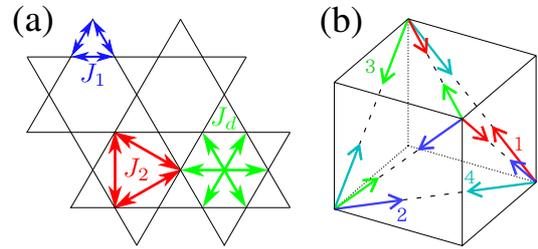}
\caption{(Color online)
(a) The nearest-neighbor couplings on the Kagome lattice are denoted as $J_1$.
The next-nearest-neighbor spins on the triangles in hexagons are coupled with $J_2$, but there is no interaction for the other triangles of the next-nearest-neighbor spins.
The 3rd-nearest-neighbor couplings $J_d$ are defined between the diagonal spins in the hexagons. 
(b) The cuboc spin configuration in the spin space. The index assigned for the surface triangles of the tetrahedron corresponds to that for the 120$^\circ$ structure defined in Fig. 2.
}
\end{figure}

The planar structure in the stacked Kagome systems is depicted in Fig. 1(a), where $J_1$ is the nearest-neighbor coupling and $J_2$ denotes the half of the next-nearest-neighbor couplings.
We also introduce the 3rd-nearest-neighbor couplings $J_d$ that run along the diagonal directions of the hexagons in the Kagome lattice.
Then, the Hamiltonian is explicitly written as
\begin{eqnarray}
{\cal H} = J_1\sum_{\langle i,j\rangle_1} {\bm S}_i\cdot {\bm S}_j +J_2\sum_{\langle i,j\rangle_2} {\bm S}_i\cdot {\bm S}_j + J_d\sum_{\langle i,j\rangle_d} {\bm S}_i\cdot {\bm S}_j  \nonumber\\
 + J_c\sum_{\langle i,j\rangle_c} {\bm S}_i\cdot {\bm S}_j
\label{model}
\end{eqnarray}
where ${\bm S}$ denotes the vector  spin of the O(3) symmetry with ${|\bm S|}=1$ and $J_c$ represents the inter-layer couplings.
We basically assume $|J_c|=1$.
Note that $J_c$ causes no frustration effect, implying that spin structures of the ordered phases are determined by the frustrating couplings in the Kagome plane.

\begin{figure}[tb]
\centering\includegraphics[width=1.0\linewidth]{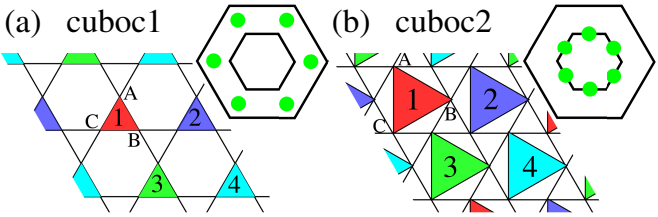}
\caption{(Color online)
(a) Cuboc1 spin structure:  the four tilting 120$^\circ$ planes at the small numbered triangles in the Kagome lattice form the tetrahedron in Fig. 1(b),  accompanying the translational symmetry breaking.
In the extended Brillouin zone corresponding to the outer hexagon in the upper right panel, the cuboc1 is represented as the modes indicated by the solid symbols.
(b) Cuboc2 spin structure, the  120$^\circ$ planes are located on the triangles of the $J_2$ couplings.
In the Brillouin zone of the inner hexagon,  the cuboc2 is represented as the solid symbols at the M points.
The numbers of the triangles correspond to those of the surface triangles in Fig. 1(b).}
\end{figure}

In order to see the formation of the cuboc orders for Eq. (\ref{model}), we consider the $J_1-J_d$ Kagome system with $J_2=0$ for a while.
As shown in Ref. \citen{Messio}, we can see that the cuboc1 is observed for $J_1>0$ and $J_d >0$,  where the $J_d$ couplings select the cuboc1 state from the highly-degenerating ground states of the Kagome system having only the nearest-neighboring $J_1$ couplings.
Figure 2(a) illustrates the spin configurations of the cuboc1 order in the real space,  where the $J_d$ couplings stabilize the staggered spin arrays along the diagonal directions.
In order to reduce the energy due to the frustrating $J_1$ couplings on the small colored triangles in Fig. 2(a),  the spin orientations in the three staggered arrays relatively tilt to form the 120$^\circ$ structure, accompanying the translational symmetry breaking.
Then, the four tilting 120$^\circ$ planes on the small triangles labeled by 1 to 4 in Fig. 2(a) can be mapped into the tetrahedron in Fig. 1(b).
As shown in the extended Brillouin zone in the upper right panel of Fig. 2(a),  the cuboc1 configuration is characterized by the triple ${\bm q}$ located at the internally dividing point in the ratio 3:1 between the $\Gamma$ and K  points.\cite{Messio}

In the region $-2J_d <J_1< 0$ with $J_d >0$, on the other hand,  we find that the cuboc2 is realized as the ground state. 
Although $J_1$ is ferromagnetic in this case, the three staggered spin arrays along $J_d$ directions are frustrating on the hexagons and the spins on the large colored triangles in Fig. 2(b) form the 120$^\circ$ structure.
Then, the four 120$^\circ$ planes labeled by 1 to 4 in Fig. 2(b) form the cuboc2 order accompanying translational symmetry breaking, which is represented as the same tetrahedron in Fig. 1(b).
The difference between the cuboc1 and the cuboc2 is in the real-space distribution of the 120$^\circ$ planes.
In contrast to the cuboc1 case,  the cuboc2 is described by the triple ${\bm q}$ corresponding to the M points in the Brillouin zone of the Kagome lattice, as shown in the upper right panel of Fig. 2(b).
In this sense, it is an interesting problem to check the universality of the two types of cuboc transitions.

Another interesting point on the above cuboc2 phase is that, if the $J_2$ coupling turns on and $J_d$ decreases,  it can be adiabatically connected to that in the Kagome-triangular spin system,  where the positive $J_2$ coupling strongly stabilizes the 120$^\circ$ structure on the $J_2$ triangles\cite{Seki,Ochiai}. 
Moreover, the phase transitions in the Kagome-triangular system were intensively investigated in the context of the coupled spin tubes by MC simulations up to about $10^5$ spins\cite{Seki}, where the crossover depending on the ratio $|J_1|/J_2$ between the weak-1st-order and 2nd-order transitions was reported.
However, the previous MC results still contained a significant finite-size effect, so that a large scale MC simulation was required for verifying the universality of the cuboc2 transition.

In order to discuss the phase transitions for Eq. (\ref{model}), we set up three order parameters characterizing the cuboc orders. 
The sublattice magnetization $\bm M$ is defined as the expectation value of spins respectively for the 12 sublattices.
The sublattice vector spin chirality ${\bm \kappa}_V$ is defined for each triangle having the 120$^\circ$ spin structure as 
$ {\bm \kappa}_V \equiv\frac{2}{3\sqrt 3}\left[\bm S_{A}\times \bm S_B+\bm S_{B}\times \bm S_{C} +\bm S_{C}\times \bm S_{A} \right] $, where $A$, $B$, and $C$ are the index of the spins in the 120$^\circ$-structure triangle. 
As seen in Fig. 2,  the four triangles are contained in the magnetic unit cell having 12 sublattices.
Also, the sublattice scalar spin chirality $\kappa_S$, which detects the non-coplanar spin structure, is defined as
$
\kappa_S \equiv \bm S_{A} \cdot [\bm S_{B} \times \bm S_{C}],
$
where $\bm S_{A}$, $\bm S_{B}$ and $\bm S_{C}$ should belong to the three adjacent surface triangles of the tetrahedron in Fig. 1(b).
For the cuboc1, $\kappa_S$ is defined on the large triangles embedded in the hexagons in Fig. 2(a), while for the cuboc2, it is located on the small triangles in Fig. 2(b).
The alternating locations of the triangles of ${\bm \kappa}_V$ and $\kappa_S$  can be viewed as a duality relation between the cuboc1 and cuboc2.

In order to analyze the finite temperature transitions for the cuboc orders in the stacked Kagome Heisenberg systems, we employ the event-chain algorithm \cite{Bernard,Michel1, Michel2, Nishikawa} combined with the parallel tempering\cite{exchange}, which enable us to achieve the relaxation to the thermal equilibrium even for the highly frustrated systems.
We then calculate the susceptibilities of the order parameters,
$\chi_M = {\langle|\bm M|^2\rangle}/T$,
and $\chi_{V} = {\langle|\bm \kappa_V|^2\rangle}/T$,
where $T$ is a temperature.\cite{chi_s}
In actual simulations, 
the typical number of the sampling is about $10^5$ after the initial relaxation of $10^4$ MC steps, and the maximum system size is $L=72$, where $L$ is the linear dimension of the stacked Kagome system in the unit of the spin triangle (unit cell of the Kagome lattice).
Note that we terminate a single event-chain update for a spin rotation axis randomly chosen from the $x, y, z$ directions, if the total rotation angle reaches $6\pi L^3/100$.

\begin{figure}[tb]
\centering\includegraphics[width=1.0\linewidth]{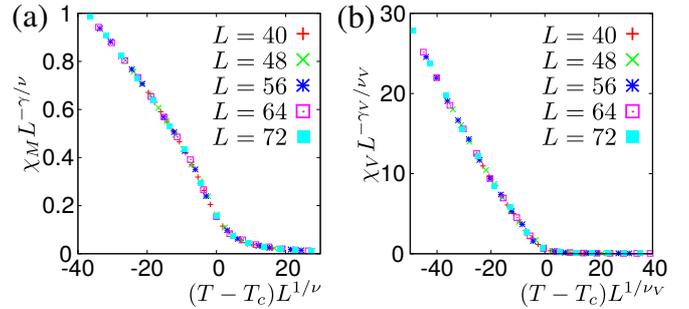}
\caption{(Color online)
Finite-size-scaling plots for the cuboc1 transition at $J_1=1, J_d=3.2, J_c=-1$.
(a) The plot of $\chi_M$, where the exponents are estimated as $\nu=0.63(2)$ and $\gamma=1.22(5)$ with  $T_c=2.1549(3)$.
(b) The plot of  $\chi_V$, where the exponents are estimated as $\nu_V=0.60(2)$ and $\gamma_V=0.53(4)$.
The error bars of the data are negligible compared with the symbols.
}
\label{FSS_cuboc1}
\end{figure}

\begin{figure}[bt] 
\centering\includegraphics[width=1.0\linewidth]{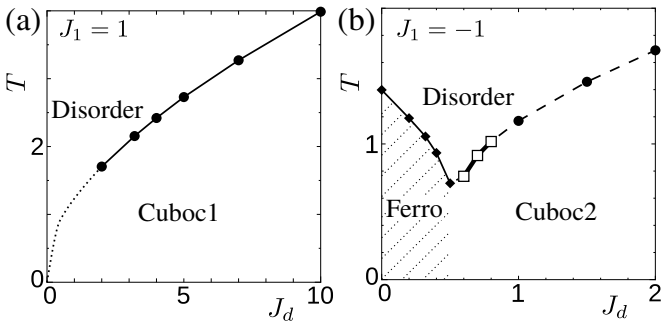}
\caption{(Color online)
Phase diagrams of the $J_1-J_d$ Kagome spin systems with $J_c=-1$ in the $T-J_d$ plane.
(a) Antiferromagnetic coupling $J_1=1$:  the 2nd-order-transition behaviors to the cuboc1 are observed in wide range of $J_d$. 
(b) Ferromagnetic coupling $J_1=-1$: 
Open squares indicate the 1st-order transition to the cuboc2 order observed in $0.5 \lesssim  J_d \le 0.8$.
For $ J_d \gtrsim 0.8$, the 2nd-order-transition behaviors are observed for $\chi_M$ within $L=72$, but the critical exponents extracted are non universal. 
}
\label{j1-jd_phase}
\end{figure}

Let us first discuss the cuboc1 transition for the $J_1-J_d$ stacked Kagome system with $J_2=0$ and $J_c=-1$ fixed.
In $J_1>0$ and $J_d>0$, the ground state has the cuboc1 order.
We then show the finite-size-scaling (FSS) plot of $\chi_M$ up to $L=72$ for $J_1=1$,$ J_d=3.2$  and  $J_c=-1$  in Fig. \ref{FSS_cuboc1}(a), which basically indicates the 2nd-order transition.
Indeed, the transition temperature and the critical exponents are extracted as  $T_c=2.1549(3)$, $\nu=0.63(2)$, and  $\gamma=1.22(5)$,  with use of the Bayesian approach\cite{Harada}.
Taking account of the $L$-dependence of the estimated exponents, we adopt  
\begin{equation}
\nu=0.60(4)\quad  {\rm and}\quad  \gamma=1.17(10)
\label{cuboc1_exp}
\end{equation} 
as the exponents for $J_1=1$, $J_d=3.2$ and $J_c=-1$.
In Fig. \ref{FSS_cuboc1}(b), we also present the FSS plot for $\chi_V$, which yields the exponents $\nu_V=0.59(2)$ and $\gamma_V=0.56(3)$ at the same $T_c$ within the numerical accuracy.\cite{chi_s}
Assuming the 2nd-order transition, moreover, we have verified that  Eq. (\ref{cuboc1_exp}) consistently reproduces the FSS plots in $3.2\le J_d\le 100$.
The results are summarized in the phase diagram of Fig. \ref{j1-jd_phase}(a).
Note that, in the region $0<J_d \lesssim 1.0$, the MC simulation fails in the relaxation to the thermal equilibrium state, due to the significant frustration originating from the $J_1$ Kagome couplings.

Although the above FSS scaling analysis basically suggests the 2nd-order transition for the cuboc1, we should remark the possibility of a weak-1st-order transition. 
This is because, for Eq. (\ref{cuboc1_exp}), the anomalous dimension $\eta\equiv 2-\gamma/\nu $ can be negative within the error bars, which could be a signature of the 1st-order transition. 
Also, the slightly large error bars in Eq. (\ref{cuboc1_exp}) originate from a non-negligible weak $J_d$ dependence involved in the FSS analysis within $L=72$, which might suggest a crossover to the 1st-order transition.
In order to establish the cuboc1 transition, we need a larger-scale simulation, which is an important future issue.

We next turn to the cuboc2 transition for the $J_1-J_d$ stacked Kagome systems in the ferromagnetic $J_1$ regime.
In  $0\le -J_d/J_1 \le 0.5$, the ground state is ferromagnetically ordered, while the cuboc2 order is realized for $ -J_d/J_1 > 0.5$. 
Thus, the finite-temperature transition to the cuboc2 order can be observed in  $ -J_d/J_1 \gtrsim 0.5$.
In the following, we fix $J_1=J_c=-1$ and vary $J_d$.
The phase diagram determined by the MC simulations up to $L=72$ is summarized in Fig. \ref{j1-jd_phase}(b).
In $J_d \lesssim 0.5$, we have observed the ferromagnetic transition, which is consistent with the ground-state behavior. 
Note that this ferromagnetic transition is verified to be in the 3D ferromagnetic Heisenberg universality.
For $J_d\gtrsim 0.5$, on the other hand, we find that the cuboc2 transition actually appears at a finite temperature.

\begin{figure}
\centering\includegraphics[width=1.0\linewidth]{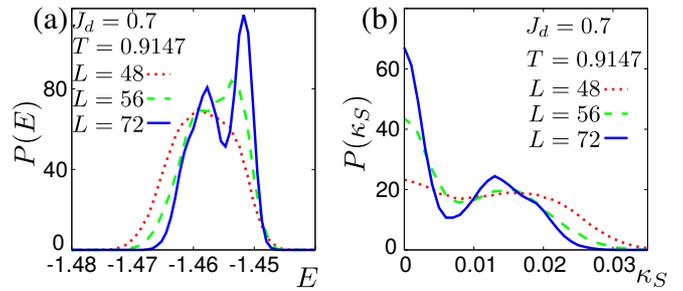}
\caption{(Color online)
Double-peak structures of histograms for the $J_1-J_d$ stacked Kagome system for $J_1=-1.0$, $J_d=0.7$, and $J_c=-1$ at $T=0.9147$. (a) The energy histogram $P(E)$ with $E$ being the energy per spin and (b) the histogram $P(\kappa_S)$ for the scalar spin chirality per triangle. 
These histograms are normalized, so that their total area is unity.
}
\label{cuboc2_dos}
\end{figure}

Let us discuss the nature of the cuboc2 transition in details.
An important point is that the 1st-order transition is confirmed in the small $J_d$ regime ($0.5 \lesssim J_d \le 0.8$).
Figure \ref{cuboc2_dos}(a) illustrates the energy histogram for $J_d=0.7$ at $T=0.9147$, where the double-peak structure emerges for $L=72$.
In Fig.\ref{cuboc2_dos}(b), we can also confirm the double peaks of the scalar-spin-chrality histogram at $\kappa_S\simeq 0$ and $0.013$.\cite{chi_peak}
Taking account of the temperature range where the double peaks are observed, we adopt   $T_c=0.9147(2)$ as a transition temperature.
As $J_d$ increases beyond $J_d \sim 0.8$,  the transition crossovers to the 2nd-order transition like behaviors.
However, the FSS plots of $\chi_M$ for  $J_d >0.8$ yield non-universal exponents for $\chi_M$ gradually gliding in $\nu\simeq 0.51 - 0.72 $ and $\gamma\simeq 0.88 - 1.38$.
Moreover, we find that the FSS plot of $\chi_V$ fails in the range $J_d>0.8$. 
On the basis of these nonuniversal behaviors, we think that the cuboc2 transition is basically described by the 1st-order transition in the bulk limit, though larger-scale simulations are required for the direct evidence of the 1st-order transition.

As mentioned before, the cuboc2 order in the $J_1-J_d$ stacked Kagome system is  adiabatically connected to that in the Kagome-triangular system with $J_1 <0 $ and $J_2 >0 $.
In the following, moreover, we assume the antiferromagnetic inter-layer coupling $J_c=1$, for the purpose of a direct comparison with the coupled-spin-tube system previously studied in Ref.\citen{Seki}, where it was reported that the 2nd-order-transition  behaviors  in $-J_1/J_2 \lesssim 0.8$  crossover to the 1st-order transition in $0.8 \lesssim -J_1/J_2 \lesssim 1.0$ within $L\le 36$.
Note that the inter-layer coupling cause no frustration and thus, the nature of the transitions is irrelevant to the sign of $J_c$. 

\begin{figure}[bt] 
\centering\includegraphics[width=1.0\linewidth]{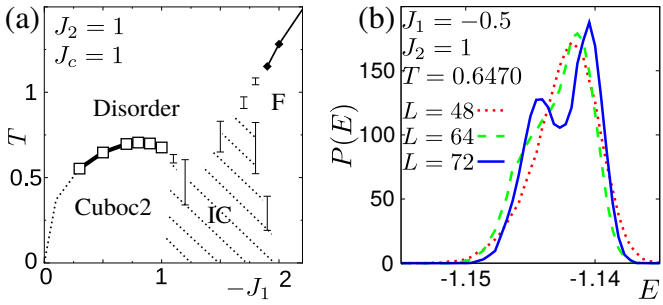}
\caption{(Color online)
(a) The $T-J_1$ phase diagram for the stacked Kagome-triangular system with $J_2=1$ and $J_c=1$. 
The open squares with the thick line indicate the 1st-order transition.
Ferromagnetic (F) and incommensurate (IC) orders are the same as Ref. \citen{Seki}.
(b) Energy histogram $P(E)$ at $T=0.6470$ for $J_1=-0.5$, which shows the double-peak structure.}
\label{tube}
\end{figure}

As $L$ increases up to $L=72$, in this paper, we have found that the FSS plot fails in extracting the critical exponents even for $-J_1/J_2=0.5$ and then the energy histogram exhibits the double-peak structure at $T_c=0.6470(1)$ estimated, as shown in Fig. 6(b).
Also we have confirmed that the histograms of the spin chiralities exhibit the double-peak structures at the same $T_c$.
These results imply that the region where the 1st-order transition appears is extended to the small  $-J_1/J_2$ side, as $L$ increases.
We then illustrate the $T-J_1$ phase diagram for the stacked Kagome-triangular system of $J_2=1$ and $J_c=1$ in Fig. \ref{tube}(a), where the 1st-order transition is confirmed in $0.3\le -J_1/J_2 \le 1.0$.
Note that  the double-peak structures of the histograms can be also confirmed in a wide range of the inter-layer coupling $J_c$ within the relatively small system sizes compared with the $J_1-J_d$ Kagome case.
A reason for this weak finite-size effect is that the $J_2$ coupling strongly stabilizes the 120$^\circ$ structure, so that the effective length scale at $T_c$ could be  shorter than that of the $J_1-J_d$ Kagome systems.

In this letter, we have investigated the phase transitions in the stacked $J_1-J_d$ Kagome Heisenberg systems, which exhibit the cuboc1 order for $J_1>0$ and the cuboc2 order for $-2J_d \lesssim J_1<0$.
We also analyzed the cuboc2 transition in the stacked Kagome-triangular spin systems.
An important point is that the cuboc1 and cuboc2 orders are represented as the triple-${\bm q}$ structures in the momentum space, though they have similar but distinct real-space configurations of the non-coplanar spin structure.
From the effective field theoretical view point, then, the phase transitions are basically described by the O(3) $\times$ O(3) LGW theory\cite{Kawamura_1990, Kawamura_1998}, for which the 1st-order transition is expected on the basis of the $4-\varepsilon$ expansion\cite{Calabrese}.
In this sense, the 1st-order transition for the cuboc2 order, which is supported by the double-peak structure of the energy histogram at the transition temperature,  is consistent with the field theoretical analysis.
Although the 2nd-order-like behaviors are observed for the large $J_d$ region of the $J_1-J_d$ Kagome system within $L=72$, we think that the  nonuniversal exponents extracted is a signature of the 1st-order transition.

In contrast,  the FSS analysis for the cuboc1 transition suggests the 2nd-order transition behaviors within $L=72$, which might contradict to the analysis of the O(3) $\times$ O(3) LGW theory.
However, the resulting anomalous dimension $\eta=2-\gamma/\nu $ estimated for the cuboc1 transition can be negative within the error bars,  which could also be  a signature of the 1st-order transition. 
We think that the discrepancy of the finite-size effects between the cuboc1 and cuboc2 transitions are basically attributed to the nonuniversal features depending on the model parameters.
For the Kagome-triangular cases where $J_2$ coupling strongly stabilizes the 120$^\circ$ structure embedded in the cuboc2 order, we have actually observed the clear evidence of the 1st-order transition.
In order to confirm the nature of the cuboc1 transition, larger scale numerical simulations are clearly required.
Also,  it is an important problem to investigate the universality/nonuniversality for other non-coplanar spin orders\cite{Reimers}, taking into account the various field-theoretical approaches\cite{Azaria, Pelissetto_2001, Pelissetto_2002, Delamotte}.

Although the phase transitions and the critical phenomena are a long standing issues in physics,  the quantitative analysis for the non-coplanar spin orders has been less explored, since the targeting systems are basically highly frustrated. 
We have demonstrated that the event-chain MC simulation is a very powerful tool for the quantitative analysis of the stacked Kagome Heisenberg systems, which are typical systems exhibiting the phase transitions to nontrivial non-coplanar spin orders.
Our results stimulate further researches of the phase transitions associated non-coplanar spin orders, which involve a new frontier of statistical mechanics and condensed matter physics.

\begin{acknowledgments}
We would like to thank T. Okubo for valuable discussions.
One of the authors(K.S) also thank K. Hukushima for discussions about the event-chain algorithm.
This work is supported by JSPS KAKENHI Grants, No. 16J02724  and  17H0931.
\end{acknowledgments}


\begin{thebibliography}{99}
\bibitem{chi_s} In general,  $|\kappa_S|^2/T$ is difficult to estimate accurately, since $\kappa_S$ is a higher order correlation with respect to the spins.
\bibitem{chi_peak} The histogram of $\kappa_S$ has the symmetric distribution in the negative $\kappa_S$ region, which is not shown in Fig. 5(b).
\end{thebibliography}
\end{document}